\documentclass[iop,superscriptaddress]{emulateapj}

\usepackage[usenames]{xcolor}

\usepackage{hyperref}

\newcommand{\mnu}{$\sum m_\nu$}
\newcommand{\Neff}{$N_\mathrm{eff}$}
\newcommand{\hh}{$H_0$}
\newcommand{\omm}{$\Omega_\mathrm{m}$}

\newcommand{\Mpc}{\textrm{Mpc}}
\newcommand{\s}{\mathrm{s}}
\newcommand{\km}{\, \mathrm{km}}

\newcommand{\eV}{\, \textrm{eV}}
\newcommand{\MeV}{\, \textrm{MeV}}

\newcommand{\cmbwmap}{CMB$_\mathrm{WMAP}$}
\newcommand{\cmbspt}{CMB$_\mathrm{SPT}$}
\newcommand{\LCDM}{$\Lambda$CDM}

\newcommand\figref[1]{%
Fig.~\ref{fig:#1}}

\newcommand\secref[1]{%
Sec.~\ref{sec:#1}}

\slugcomment{Prepared for ApJ}

\shorttitle{Number and mass of relativistic species}
\shortauthors{Riemer-S\o{}rensen et al.}

\begin{document}

\title{Simultaneous constraints on the number and mass of relativistic species}


\author{Signe Riemer-S\o{}rensen}\email{signe@physics.uq.edu.au} 
\author{David Parkinson}
\author{Tamara M. Davis}
\affiliation{School of Mathematics and Physics, University of Queensland, Brisbane, QLD 4072, Australia}
\author{Chris Blake}
\affiliation{Centre for Astrophysics \& Supercomputing, Swinburne University of Technology, P.O. Box 218, Hawthorn, VIC 3122, Australia}


\begin{abstract}
Recent indications from both particle physics and cosmology suggest the existence of more than three neutrino species. In cosmological analyses the effects of neutrino mass and number of species can in principle be disentangled for fixed cosmological parameters. However, since we do not have perfect measurements of the standard $\Lambda$ Cold Dark Matter model parameters some correlation remains between the neutrino mass and number of species, and both parameters should be included in the analysis. Combining the newest observations of several cosmological probes (cosmic microwave background, large scale structure, expansion rate) we obtain $N_\mathrm{eff}=3.58^{+0.15}_{-0.16} (68\%\mathrm{\, CL})^{+0.55}_{-0.53}(95\%\mathrm{\, CL})$ and $\sum m_\nu < 0.60\eV (95\%\mathrm{\, CL})$, which are currently the strongest constraints on \Neff{} and \mnu{} from an analysis including both parameters. The preference for \Neff{}$>3$ is now at a $2\sigma$ level.
\end{abstract}



\section{Introduction}
Neutrinos are the lightest massive known particles, and despite the fact that we know they have mass, they are treated as exactly massless by the Standard Model of particle physics. Neutrino oscillation experiments using solar, atmospheric, and reactor neutrinos have measured mass differences between the three Standard Model species\footnote{Technically they are not Standard Model neutrinos because they have mass, but here we refer to the $\nu_e$, $\nu_\mu$, and $\nu_\tau$ as Standard Model, and any additional species as sterile.} to be $\Delta m_{32}^2 = |(2.43^{+0.12}_{-0.08}) \times10^{-3}|\, \eV^2$ and $\Delta m_{21}^2 = (7.50\pm0.20)\times10^{-5}\, \eV^2$ \citep{Fukuda:1998,Beringer:2012}. The Heidelberg-Moscow experiment has limited the mass of the electron neutrino to be less than $0.35\, \eV$ using neutrino-less double $\beta-$decay \citep{Klapdor:2006}, but no current laboratory experiment has the sensitivity to measure the absolute neutrino mass. 

The oscillation data from short baseline experiments exhibit some tension allowing for, or even preferring, the existence of additional neutrino species. Depending on the exact analysis the preferred scenarios include either one or two sterile neutrinos in addition to the three normal ones (3+1 or 3+2) \citep{Kopp:2011,Mention:2011,Huber:2011,Giunti:2011}.

Observations of the Cosmic Microwave Background (CMB) also seem to favour additional species of radiation to be present at the time of decoupling over and above that provided by the photon density and three Standard Model neutrinos; most notably the results from the Atacama Cosmology Telescope (ACT) \citep{Dunkley:2011} of $N_\mathrm{eff} =  5.3 \pm 1.3$ but other authors e.g.\ \citet{Komatsu:2010,Hou:2011,Keisler:2011} find a similar preference. \Neff{} is the parametrisation in terms of the effective number neutrino species. In analogy to dark energy and dark matter, this extra radiation is often dubbed dark radiation. 

The neutrino oscillation results favour a large mass difference \citep[e.g.][]{Kopp:2011} providing a lower limit on the sterile neutrino mass of the order of $1\eV$, which is incompatible with cosmological mass constraints from combinations of CMB and Large Scale Structure (LSS) measurements \citep[$\sum m_\nu <0.3-0.6 \eV$, e.g.][]{Riemer-Sorensen:2012,Archidiacono:2012,Joudaki:2012b}, but it has been shown that factors such as initial lepton asymmetry can possibly alleviate these constraints by introducing a non-thermal neutrino velocity spectrum and thereby avoiding the LSS mass constraints \citep{Hannestad:2012}.

Many recent publications have addressed the issues of neutrino mass and effective number of neutrinos separately \citep{Moresco:2012,Xia:2012,dePutter:2012,Riemer-Sorensen:2012} but very few consider the correlation between the two. In this paper we address these correlations and demonstrate that due to imperfect measurements of the standard $\Lambda$ Cold Dark Matter (\LCDM{}) parameters, the two parameters are not entirely independent but should be addressed simultaneously. \secref{Physical} explains the physical and observable effects of \Neff{} and \mnu{}, \secref{data} and \secref{method} describe the data and methods used to obtain the results of cosmological fits which are presented in \secref{Results}.

\section{Physical and observable effects} \label{sec:Physical}
\begin{figure*}
	\includegraphics[width=\textwidth]{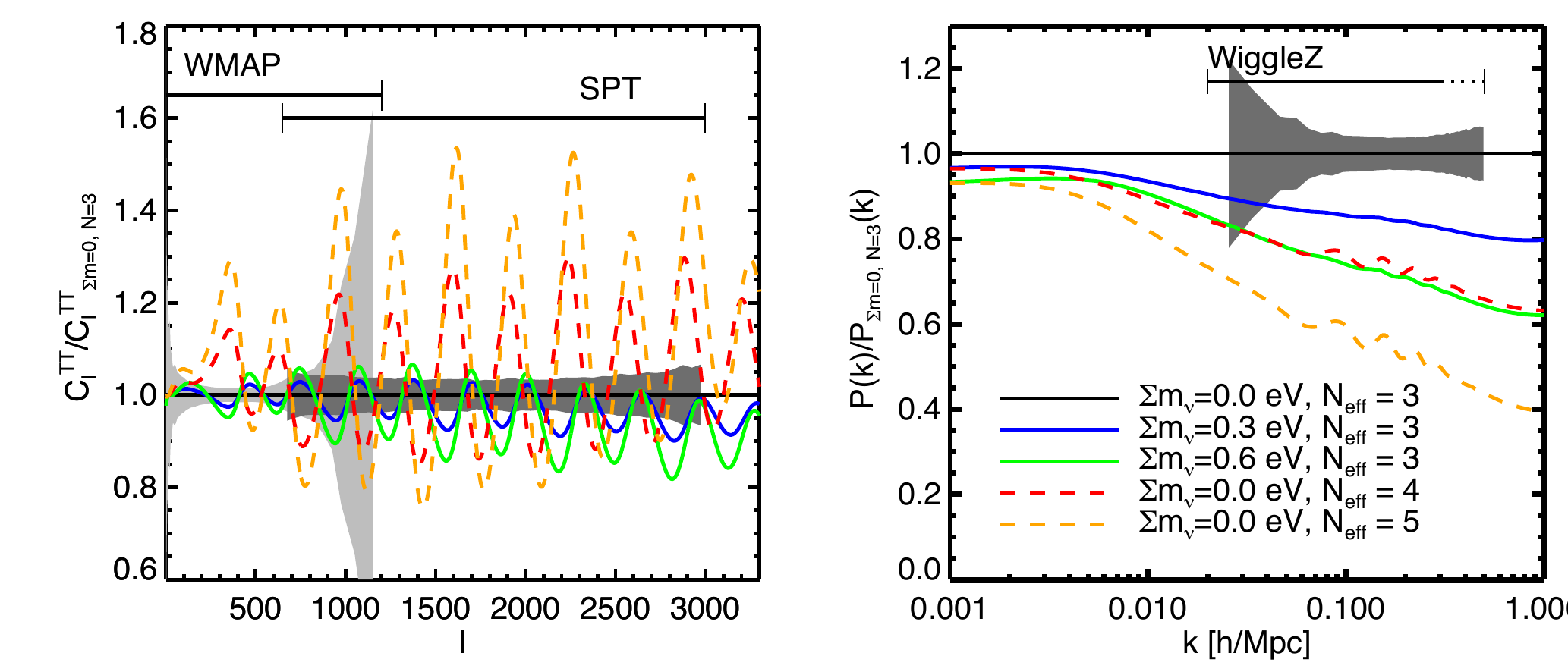} 
	\caption{Illustration of how the CMB and matter power spectra change for varying neutrino mass (solid lines) and effective number of neutrinos (dashed lines) fixing all other parameters (to WMAP 7-yr values for \LCDM).
	 \mnu{} does not affect the CMB power spectrum much, but changes the matter power spectrum significantly. The effect of \Neff{} is clearly visible for small scales (high values of $l$) in the CMB power spectrum, and the two parameters are clearly degenerate in particular for the matter power spectrum. The shaded regions indicate the normalised uncertainties of current experiments. 
	 }
	\label{fig:schematic}
\end{figure*}

This section briefly explains the physical effects of \mnu{} and \Neff{} and how they can be measured.
 
The parameter that is probed directly by cosmological analyses is not the neutrino mass, but the neutrino density, $\rho_\nu$, which can be expressed in terms of the mass \citep{Lesgourgues:2006}:
\begin{equation}
\Omega_\nu = \frac{\rho_\nu}{\rho_c} = \frac{\sum_{i=1}^{N_\nu} m_{\nu,i}}{93.14\eV h^2}
\end{equation}
where $\rho_c$ is the critical energy density for a flat Universe, and $N_\nu$ is the number of massive neutrino states. Because of the smallness of the measured mass differences relative to the upper limits, it is reasonable to assume that that if the individual neutrino masses are near the upper limit, they are effectively equal.

\subsection{Massive neutrino effects on the expansion}
Due to the weakness of the Weak interaction the neutrinos decouple early from the tightly coupled baryon-photon fluid ($T=2-4\MeV$), while the expansion is still dominated by radiation. If $\sum m_\nu >1\eV$ the neutrinos become non-relativistic before recombination. However, cosmological constraints already indicate $\sum m_\nu <1\eV$ so the neutrinos become non-relativistic after recombination, and while they are relativistic their energy density contributes as radiation rather than matter. The presence of ``neutrino radiation" affects the expansion rate and changes the time of the matter-radiation equality. The latter determines when the density perturbations can begin to collapse gravitationally and structures can evolve. The higher \mnu{} or \Neff{}, the higher the radiation density and the later the equality takes place. Changing the expansion rate and time of the matter-radiation equality has some measurable consequences described below and illustrated in \figref{schematic}:
\begin{itemize}
\item The position of the peak of the matter power spectrum is determined by the particle horizon length scale at the time of matter-radiation equality. Moving the equality to later times shifts the peak to larger scales.

\item The expansion rate determines the time available for sound waves to propagate in the baryon-photon fluid before recombination, which will imprint on the BAO scale measurable from both the CMB and the large scale structure. Increasing the expansion rate will move the BAO peak to smaller scales.

\item Delaying matter-radiation equality enhances the ratio of even to odd peaks in the CMB power spectrum because the dark matter potentials have longer time to grow, and thereby increasing the BAO amplitude (the gravitational potentials are deeper, but the photon restoring force is the same). It also increases the early Sachs-Wolfe effect, which enhances the overall normalisation of the CMB power spectrum. 

\item Changing the expansion rate can significantly affect the Big Bang Nucleosynthesis (BBN) leading to different deuterium and helium abundances than the observationally confirmed predictions of standard BBN \citep{Lesgourgues:2006,Steigman:2012}.
\end{itemize}

\subsection{Massive neutrino effects on structure formation}
When the neutrinos have decoupled, their thermal velocities decay adiabatically as \citep{Komatsu:2010}:
\begin{equation}
v_\mathrm{thermal}=151(1+z)(1\eV/m_\nu)\km\sec^{-1} \, .
\end{equation}
When they become non-relativistic they behave as a species of warm/hot dark matter, suppressing density fluctuations on scales smaller than their free-streaming length (at the time when they become non-relativistic) \citep{Lesgourgues:2006}:
\begin{eqnarray}
k_\mathrm{FS} 	&=& \sqrt{\frac{3}{2}}\frac{H(t)}{v_\mathrm{thermal}(t)(1+z)} \\
			&\approx& 0.82\frac{\sqrt{\Omega_\Lambda+\Omega_m(1+z)^3}}{(1+z)^2}\left(\frac{m_\nu}{1\eV}\right)h \Mpc^{-1} \, .
\end{eqnarray}

Neutrinos cannot cluster on scales smaller than their free-streaming length and consequently the density fluctuations are suppressed by a factor proportional to the neutrino density. This is observable in the density power spectrum as a suppression of structure on small scales (see left part of \figref{schematic}). However, the presence of neutrinos also enhances the density perturbations (by adding to the radiation density for fixed \omm{}), which boosts structure formation in general, so for higher \Neff{}, $\sigma_8$ \citep{Hou:2011} will increase.

The evolution of the density perturbations can be predicted by solving the coupled set of Einstein and Boltzman equations. For small density perturbations the linearised equations are sufficient (e.g. using CAMB\footnote{\url{camb.info}}), but when the density perturbations grow sufficiently large, the higher-order, non-linear terms become important and have to be included, which is computationally very intensive (and not yet solved analytically). Instead one has to rely on second order perturbation theory \citep[e.g.][]{Saito:2009,Taruya:2012} or simulations \citep[e.g.][]{Smith:2003,Jennings:2010}.

Numerical solutions show that for $f_\nu=\Omega_\nu/\Omega_m<0.07$ the suppression is $\delta P/P =-8f_\nu$ for linear structure formation \citep{Hu:1998} and the effect increases for non-linear structure formation \citep[$8\rightarrow9.6$][]{Brandbyge:2008, Brandbyge:2009, Brandbyge:2010, Viel:2010, Agarwal:2011}.

The signature of free-streaming massive neutrinos can be detected in both the power spectrum of CMB temperature anisotropies and in the (matter) density power spectrum of large-scale structure. As illustrated in \figref{schematic}, the larger the neutrino mass, the more prominent these effects on the power spectra of the CMB and LSS.

The density power spectrum can be traced either through galaxies (galaxy power spectrum) or through the absorption lines from neutral hydrogen clouds along the line of sight to distant quasars (Lyman-alpha forest power spectrum). The neutrinos affect the total matter power spectrum, which may differ slightly from the tracer power spectrum that we actually measure. Depending on the tracer, the bias, $P_\mathrm{obs}=b^2P_\mathrm{m}$, might be scale dependent or scale-independent \citep{Schulz:2006}. 

Another effect that is important for spectroscopic galaxy surveys is the redshift space distortions caused by galaxies falling into the gravitational potential of galaxy clusters \citep{Kaiser:1987,Percival:2009b}. As structure formation becomes non-linear, so do the redshift space distortions \citep{Jennings:2010,Marulli:2011}. 

\section{Data} \label{sec:data}
For the analysis presented here we have used the following publicly available data sets:

\begin{description}
  \item[\cmbwmap] The CMB as observed by Wilkinson Microwave Anisotropy Probe (WMAP) from the 7 year data release\footnote{Available from \url{http://lambda.gsfc.nasa.gov}} \citep{Komatsu:2010}.
  
  \item[\cmbspt] The CMB as observed by the South Pole Telescope\footnote{Available from \url{http://pole.uchicago.edu/public/data/keisler11/}} (SPT) \citep{Keisler:2011}.
  
  \item[\hh{}] A Gaussian prior of $H_0 = 73.8\pm2.4\, \km \, \s^{-1} \, \Mpc^{-1}$ on the Hubble parameter value today from \citet{Riess:2011}.

  \item[$H(z)$] The expansion history of the Universe as measured with passively evolving galaxies at $z<1.75$. The combined sample of \citet{Simon:2005,Stern:2010,Moresco:2012b} as given in \citet{Moresco:2012}. The expansion history is consistent with $H(z)$ as measured from the Alcock-Paczynski test with WiggleZ galaxies \citep{Blake:2012}.
  
  \item[BAO] The BAO scale measured at $z=0.106$ ($r_s/D_V = 0.336\pm0.015$) from the Six Degree Field Galaxy Survey (6dFGS) \citep{Beutler:2011}, the reconstructed value at $z=0.35$ ($r_s/D_V = 0.1126\pm0.0022$) from Sloan Digital Sky Survey (SDSS) Luminous Red Galaxies (LRG) \citep{Padmanabhan:2012}, and $z=0.57$ ($r_s/D_V = 0.0732\pm0.0012$) from Baryon Oscillation Spectroscopic Survey (BOSS) \citep{Anderson:2012}. The measurements of $r_s/D_V$ are independent of the assumed fiducial values of \Neff{} (as they cancel out).
  
  \item[SN] Type Ia supernovae as observed by the SuperNova Legacy Survey (SNLS) \citep{Guy:2010,Sullivan:2011,Conley:2011}.
  
  \item[P(k)] The WiggleZ Dark Energy Survey power spectrum\footnote{Available from \url{http://smp.uq.edu.au/wigglez-data}} as described in \citet{Parkinson:2012}.
\end{description}

\subsection{Comments on data combinations}
WMAP and SPT are highly complementary observations of the CMB. WMAP is a full sky survey including the largest scales (multipole moments $l=2-1200$) while SPT is a high resolution small scale survey ($l=650-2999$) very sensitive to effects of changing \Neff{} (see the right panel of \figref{schematic}).

The large scale structure can be included either in the form of the BAO scale or the full power spectrum shape. To be conservative we do not include both the BAO distance measurement and the power spectrum shape data from the same survey since we do not know the correlation between the two. For our main result we use the BAO scale from 6dFGS, SDSS and BOSS, and the full WiggleZ power spectrum. We have tested that the results from this setup are very similar to using the full power spectrum shape from both SDSS and WiggleZ. This is because the constraints are limited by the determination of the \LCDM{} parameters rather than by the observational determination of the slope of the galaxy power spectrum. The precise determination of the BAO scale strengthens the constraints on the \LCDM{} parameters, which improves the \mnu{} and \Neff{} constraints by breaking degeneracies. Therefore including some BAO data is more important than including the full power spectrum shape for all LSS data.

A recent analysis based re-calibration of the cepheids found $H_0 = 74.3\pm2.1\,\mathrm{km\,s^{-1}\,Mpc^{-1}}$ \citep{Freedman:2012} which is slightly larger than the value adopted here driving the value of \Neff{} upwards. However, the tightest constraints presented here does not rely on the \hh{} measurement but the BAO and SN to constrain the expansion history.

For the SN we use the SNLS data which is a consistent sample of SN observed with a single telescope and consequently free from some of the ligthcurve fitting issues that \citet{Joudaki:2012b} pointed out between the Union2 and SDSS SN data sets.

\section{Method} \label{sec:method}
We fit the data to a standard flat $\Lambda$CDM cosmology with the following parameters: the physical baryon density ($\Omega_\mathrm{b}h^2$), the physical cold dark matter density ($\Omega_\mathrm{dm}h^2$) the Hubble parameter at $z=0$ ($H_0$) the optical depth to reionisation ($\tau$), the amplitude of the primordial density fluctuations ($A_s$), and the primordial power spectrum index ($n_s$). In addition we vary the effective number of relativistic degrees of freedom ($N_\mathrm{eff}=N_\nu+\Delta N_\mathrm{eff}$) and the sum of neutrino masses ($\sum m_\nu = N_\nu m_\nu$) where $N_\nu$ is the number of massive neutrinos usually taken to be $N_\nu=3.046$ and the 0.046 accounts for the neutrino energies that arise due to the residual heating provided by the $e^+e^-$-annihilations after the neutrino decoupling \citep{Mangano:2005}.

We sample the parameter space using the Markov Chain Monte Carlo sampler\footnote{http://cosmologist.info/} CosmoMC \citep{Lewis:2002}. Details on the specific modules can be found in \citet{Parkinson:2012} and on the WiggleZ website$^5$.

The probability distributions are calculated by marginalising over the density of points in the chains. Because the distributions can be non-Gaussian, the $2\sigma$ uncertainties can be larger than the $2\times1\sigma$ uncertainties.

\section{Results and discussion} \label{sec:Results}
\begin{figure}
	\includegraphics[width=8.6cm]{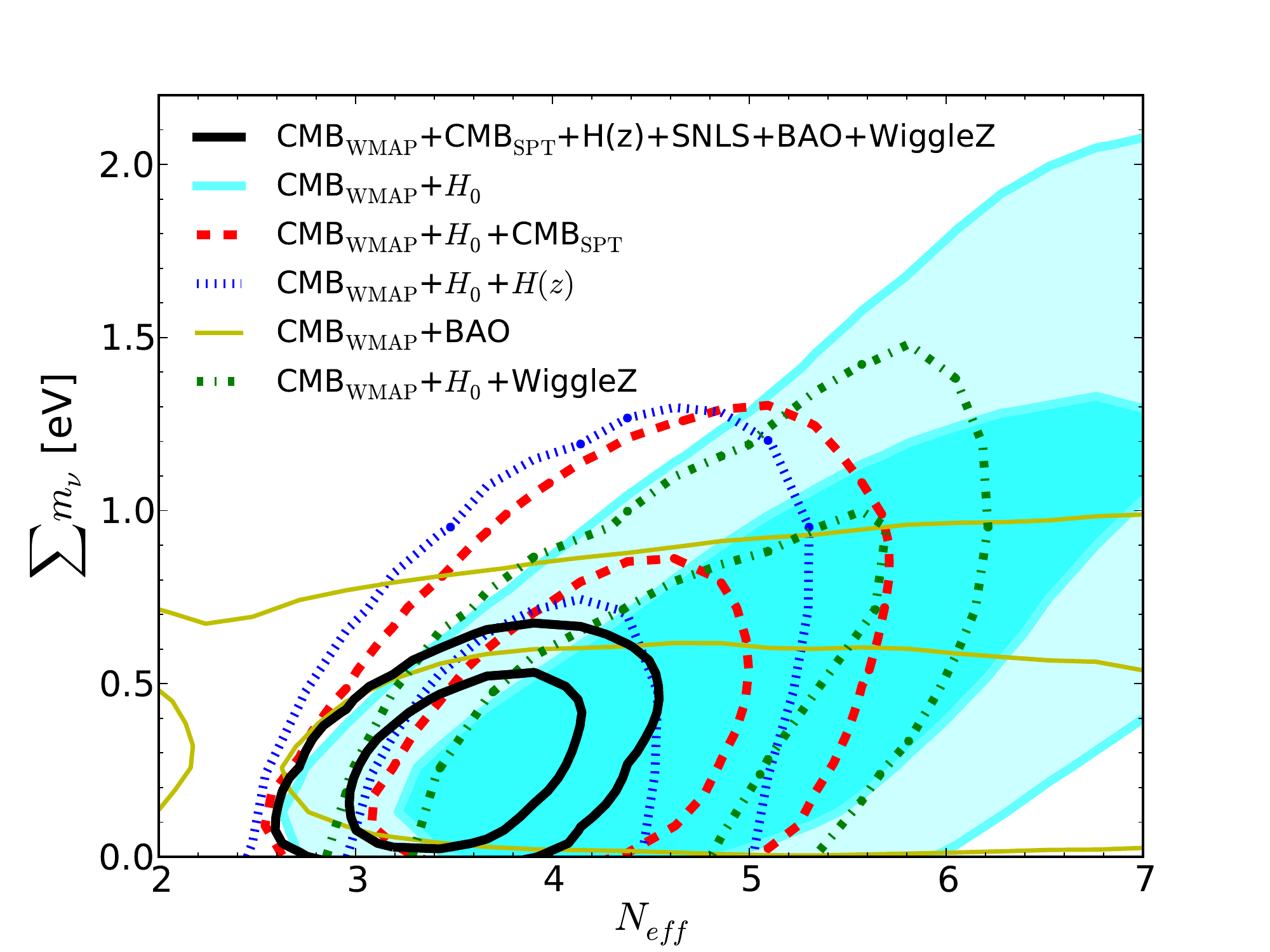} 
	\caption{The 68\% and 95\% CL contours in the \Neff-\mnu{} parameter space of fitting a \LCDM+\mnu+\Neff{} model to a selection of data combinations.}
	\label{fig:mnuNeff}
\end{figure}

\begin{figure}
	\includegraphics[width=8.6cm]{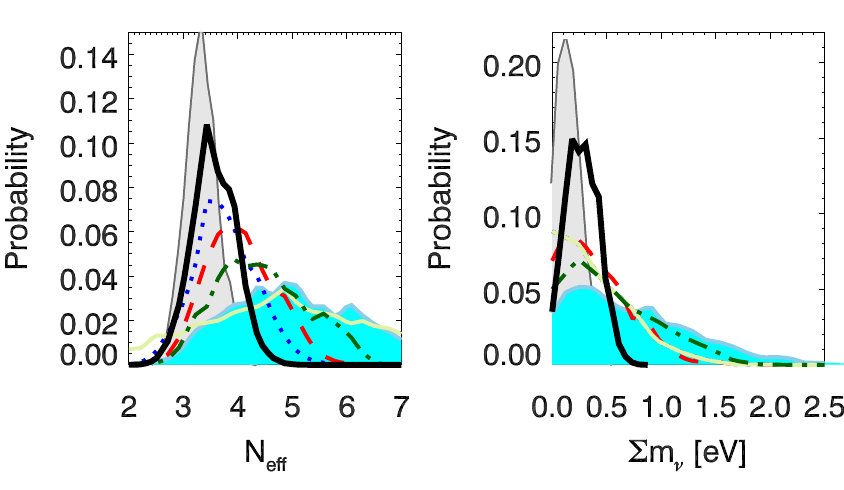} 
	\caption{The one dimensional probability distributions for \Neff{} and \mnu{} of fitting a \LCDM+\mnu+\Neff{} model to a selection of data combinations (same colours as in \figref{mnuNeff}). The grey shaded area indicates the probability distribution when correlations are neglected by fixing \mnu=0 (left) and \Neff=3.046 (right).}
	\label{fig:1D}
\end{figure}

From fitting \LCDM+\mnu+\Neff{} to \cmbwmap{} + \cmbspt{} + WiggleZ + $H(z)$ + BAO + SN we obtain $N_\mathrm{eff}=3.58^{+0.15}_{-0.16} (68\%\mathrm{\, CL})^{+0.55}_{-0.53}(95\%\mathrm{\, CL})$ and $\sum m_\nu < 0.60\eV (95\%\mathrm{\, CL})$ shown in \figref{mnuNeff} and \figref{1D}. This not represents a preference for more than three neutrino species at 95\% confidence.

\subsection{Comparison to $\Lambda$CDM}
\begin{figure}
\centering
	\includegraphics[width=0.49\textwidth]{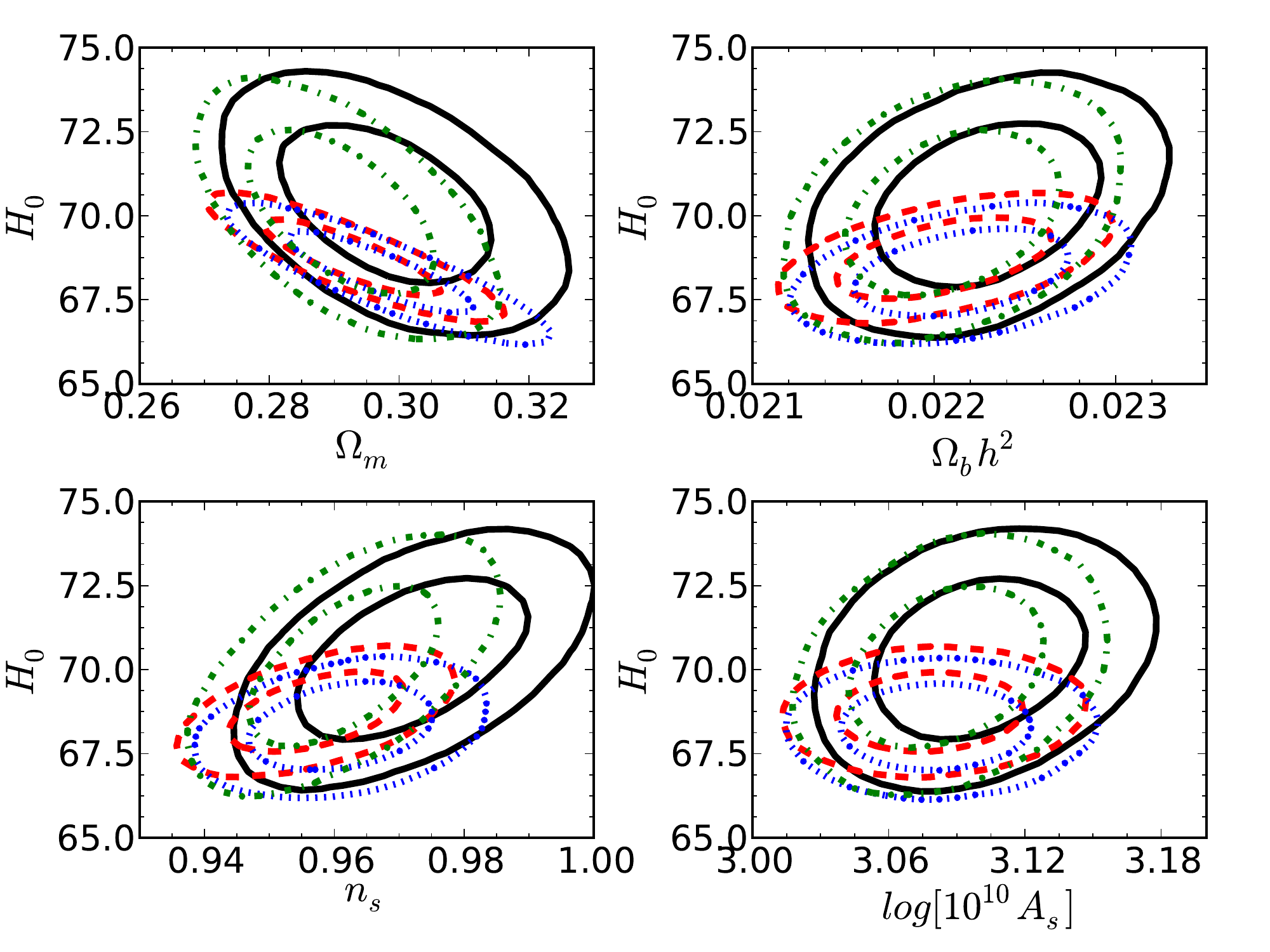} 
	\caption{The 68\% and 95\% CL contours for fitting \LCDM+\mnu+\Neff{} (solid black), \LCDM+\mnu{} (dotted blue), \LCDM+\Neff{} (dot-dashed green), and \LCDM{} (dashed red) models to \cmbwmap{} + \cmbspt{} + WiggleZ + $H(z)$ + BAO + SN data. The resulting contours are consistent with each other for all parameters.}
	\label{fig:LCDM}
\end{figure}

\figref{LCDM} shows the resulting contours from fitting \LCDM+\mnu+\Neff{} to \cmbwmap{} + \cmbspt{} + WiggleZ + $H(z)$ + BAO + SN (solid black), compared to fitting a pure \LCDM{} model (dashed red), \LCDM{}+\mnu{} (dotted blue), and \LCDM{}+\Neff{} (dot-dashed green) to the same data. Naturally the constraints are tighter for pure \LCDM, since there are fewer parameters to constrain, but overall the constraints are consistent for all parameters. Adding \mnu{} to the fit does not change the \LCDM{} parameters or uncertainties significantly whereas adding \Neff{} increases the values and uncertainties of $H_0$. Adding both parameters increases the preferred values of $H_0$ and $n_s$ but also the uncertainties, so the values remain consistent with the pure \LCDM{} case. The shift can be understood as follows: Increasing $H_0$ changes the height of the peaks in the CMB, as it corresponds to increasing the physical matter density. Changing \Neff{} recovers the details of the CMB peaks (because the original ratio of matter to radiation is recovered, restoring the details of Silk damping), but with too much power on large scales due to ISW. This shifts the primordial power spectrum from very red ($n_s=0.96$) to slightly less red ($0.98-1.0$).

\subsection{Combinations and degeneracies}
\begin{figure}
\centering
	\includegraphics[width=8.6cm]{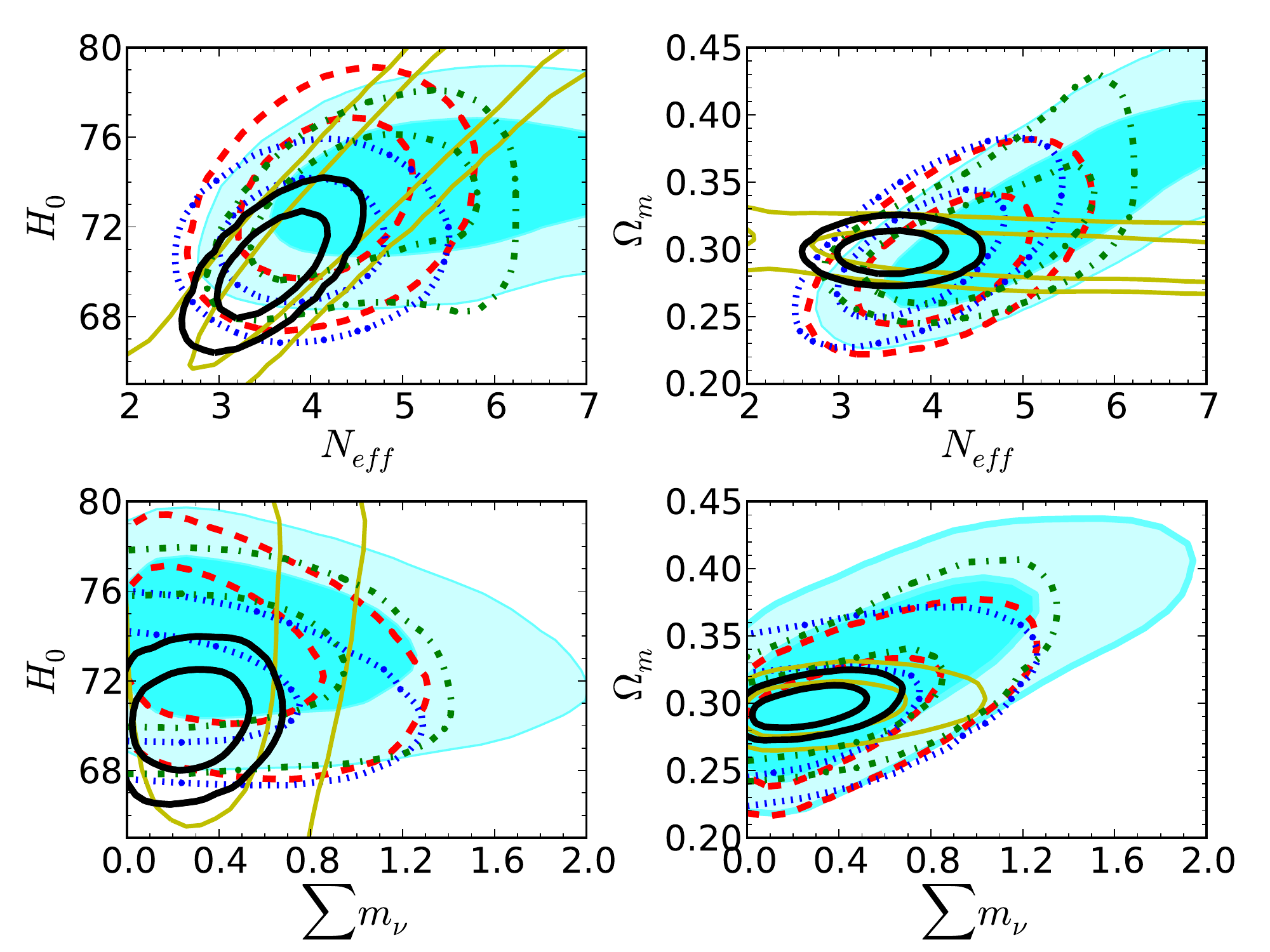} 
	\caption{Results of fitting a \LCDM+\mnu+\Neff{} model to a selection of data combinations (same colours as in \figref{mnuNeff}).}
	\label{fig:2Dplots}
\end{figure}

It is clear from the \mnu{}-\Neff{} contours in \figref{mnuNeff} that unless all data sets are included, the two parameters are correlated. The tightest constraints come from the combination of \cmbwmap{} + \cmbspt{} + WiggleZ + $H(z)$ + BAO + SN. \figref{2Dplots} demonstrates the degeneracies between \mnu{} \& \Neff{}, and \hh{} \& \omm{}. When all the data is combined, there is a significant correlation between \hh{} and \Neff{}, and a mild correlation between \mnu{} and \omm{}. Lowering \hh{} shifts \Neff{} back towards the expected value of three.

At the present, large scale structure (e.g.\ WiggleZ and SDSS) does not constrain \Neff{} uniquely because the turnover of the power spectrum is not well measured, which leaves some correlation between \omm{}, \mnu{}, and \Neff{} (see \figref{schematic}), but inclusion of the power spectrum improves on \mnu{} from the small scale suppression. 

The $H(z)$ data does not add much to the neutrino mass constraints, but because it constrains the expansion history, it improves on the limits on \Neff{} as discussed in \citet{Moresco:2012}.

\subsection{Nucleosynthesis}
There have been a number of recent attempts to derive \Neff{} from BBN alone \citep[e.g.][]{Mangano:2011, Pettini:2012} which all seem to be consistent with $N_\mathrm{eff} = 3$. Analyses combining BBN and CMB seem to prefer $N>3$ \citep[e.g.][]{Nollett:2011,Hamann:2011}. It is concerning that the preference for \Neff$>3$ is present in all analyses including CMB, but not preferred by BBN alone. This could indicate a systematic error in one of the data sets. However, BBN alone relies on very few and notoriously difficult measurements of the deuterium and helium fractions. 

The difference between the results could also be interpreted as potential tension between \Neff{} measured at two different epochs (BBN and recombination), where we naively would expect \Neff{} to be the same. A temporal variation can be explained theoretically by a decaying particle \citep{Boehm:2012} but currently there is no experimental evidence for the existence of such particle.

\subsection{Beyond \LCDM+\mnu+\Neff{}}
Other physical effects such as curvature, varying equation of state, running of the spectral index etc.\ could mimic \Neff{}$>3$ if not properly accounted for in the modelling. \citet{Joudaki:2012} demonstrated that the deviation from $N_\mathrm{eff}=3.046$ is diminished if allowing for curvature, varying equation of state, running of the spectral index, and/or the helium fraction by fitting to \cmbwmap, \cmbspt, BAO scale from SDSS and 2dFGS, $H_0$ from HST, and SN from Union2. However, for all parameter combinations the preferred value of \Neff{} was still above three, and only when more than one extra extension of the $\Lambda$CDM cosmology (e.g. curvature {\it and} varying equation of state in addition to \Neff{} and \mnu{}) were considered, did the preferred value of \Neff{} become consistent with three within one standard deviation. We take this as an indication that one extension alone does not explain the preference for $N_\mathrm{eff}\neq3$.

\subsection{Future}
Measuring the position of the peak of the matter power spectrum (the turnover) would give another handle on \Neff{}. \citet{Poole:2012b} predicts that a Euclid-like galaxy survey will be able to constrain \Neff{} to approx. 20\% independently of the CMB. However, the position of the turnover can be degenerate with neutrino hierarchy effects \citep{Wagner:2012}. The degeneracies between \mnu{}, \Neff, and the hierarchy allows for a measurement of either of them if the remaining parameters can be ``fixed'' by independent data.

If \Neff{} can be measured with sufficiently high precision, eventually it will be possible to measure the thermal distortion of the neutrino spectrum (the 0.046).

Any neutrino-like behaving particle, including sterile neutrinos and axions etc., which decouple early when relativistic and become non-relativistic, can mimic the effect of the neutrinos. If the value of $N_\mathrm{eff}>3$ is due to a particle, the result points towards physics beyond the Standard Model, and its existence it will have to be confirmed by a particle physics laboratory experiment. If no sterile neutrinos are found in laboratory experiments, the cosmological preference for additional species may indicate a lack of understanding of early Universe physics.

\section{Conclusion}
Due to imperfect measurements of the \LCDM{} parameters \mnu{} and \Neff{} are not entirely independent parameters and should be fitted simultaneously in cosmological analyses. Performing such fit of \LCDM+\mnu+\Neff{} to a combination of cosmological data sets, leads to a $2\sigma$ preference for \Neff$>3.58^{+0.55}_{-0.53} (95\% \mathrm{(CL)}$ and $\sum m_\nu <0.60\eV$, which are currently the strongest constraints on \Neff{} from cosmology simultaneously fitting for \Neff{} and \mnu{}.

\bibliographystyle{apj}
\bibliography{mnuNeff}

\end{document}